\documentclass[a4paper,preprint]{emulateapj}
\usepackage{amstext}
\usepackage{apjfonts}
\usepackage{amsmath}

\begin{document}

\title{Implications of the Measured Image Size for the Radio Afterglow of
GRB 030329}

\author{Jonathan Granot\altaffilmark{1},
Enrico Ramirez-Ruiz\altaffilmark{1,}\altaffilmark{2} and Abraham
Loeb\altaffilmark{3}}

\altaffiltext{1}{Institute for Advanced Study, Einstein Drive,
Princeton, NJ 08540} \altaffiltext{2}{Chandra Fellow}
\altaffiltext{3}{Harvard-Smithsonian Center for Astrophysics, 60
Garden St., Cambridge, MA 02138}

\begin{abstract}

We use data on the image size of the radio afterglow of GRB 030329 (Taylor
et al. 2004) to constrain the physical parameters of this
explosion. Together with the observed broad band spectrum, this data
over-constrains the physical parameters, thus enabling to test different
GRB jet models for consistency. We consider two extreme models for the
lateral spreading of the jet: {\bf model 1} with relativistic expansion in
the local rest frame, and {\bf model 2} with little lateral expansion as
long as the jet is highly relativistic. We find that both models are
consistent with the data for a uniform external medium, while for a stellar
wind environment {\bf model 1} is consistent with the data but {\bf model
2} is disfavored by the data.  Our derivations can be used to place tighter
constraints on the dynamics and structure of GRB jets in future afterglows,
following a denser monitoring campaign for the temporal evolution of their
image size.

\end{abstract}


\section{Introduction}
\label{int}

It has long been recognized that direct imaging of Gamma-Ray
Bursts (GRBs) can provide important constraints on their physical
parameters \citep{Waxman97,Sari98,PM98,GPS99a,GPS99b,GrL01}.
Unfortunately, the characteristic size of a GRB image is only of
order a micro-arcsecond about a day after the GRB at the Hubble
distance, and so it cannot be resolved by existing telescopes.
Nevertheless, indirect constraints on the image size of GRB
afterglows were derived based on the transition between
diffractive and refractive scintillations \citep{Goodman97} in the
radio afterglow of GRB 970508 \citep{Frail97,WKF98}, and based on
microlensing by a star in a foreground galaxy \citep{LP98} for the
optical lightcurve of GRB 000301C
\citep{GLS00,GrL01,GaL01,ML01,GGL01}.

Obviously the challenge of imaging a GRB is made easier for nearby sources
where the late radio afterglow extends over a wide, possibly resolvable
angle \citep{WL99,Cen99,WL01,Pa01,GL03}.  Recently, \citet{Taylor04} have
used a VLBI campaign to measure, for the first time, the angular size and
proper motion of the radio afterglow image of the bright, nearby
($z=0.1685$) GRB 030329. The diameter of the afterglow image was observed
to be $\sim 0.07\;$mas (0.2 pc) after 25 days and $0.17\;$mas ($0.5\;$pc)
after 83 days, indicating an average velocity of $\sim 4.1-5.7\;$c. This
superluminal expansion is consistent with expectations of the standard
relativistic jet model \citep{ONP04}.  The projected proper motion of GRB
030329 was measured to be smaller than 0.3 mas for 80 days following the
GRB.

Here we use the data of \citet{Taylor04} to constrain the physical
parameters of GRB 030329 based on detailed modelling of the
collimated relativistic hydrodynamics of GRB afterglows.  Since
the current state-of-the-art modelling of afterglow jets is still
flawed with uncertainties
\citep{Rhoads99,SPH99,Granot01,KG03,SAL03,CGV04}, we use this data
to critically assess some classes of models that were proposed in
the literature.  An important difference between relativistic
radio jets of GRBs and the better-studied relativistic radio jets
of quasars \citep{BBR84} or micro-quasars \citep{MR99} is that
active quasars often inject energy over extended periods of time
into the jet while GRB sources are impulsive. Although quasar jets
remain highly collimated throughout their lifetimes, GRB jets
decelerate and expand significantly once they become
non-relativistic, $\sim 1\;$yr after the explosion. The
hydrodynamic remnant of a GRB eventually becomes nearly spherical
only after $\sim 5\times 10^3\;$yr \citep{AP01}.

The outline of the paper is as follows.  In \S \ref{self_abs}, we
discuss the expected image size of radio afterglows and its
relation to the observed flux density below the self absorption
frequency.  In \S \ref{temp_ev}, we analyze the expected temporal
evolution of the afterglow image size.  The expected linear
polarization is discussed in \S \ref{pol}, while the surface
brightness profile across the image and its effects on the
estimated source size are considered in \S \ref{SB}. Finally, we
apply these derivations to the radio data of GRB 030329 (\S
\ref{application}) and infer the physical parameters from the
measured spectrum (\S \ref{phys_par}). We conclude in \S \ref{dis}
with a discussion of our primary results and their implications.

\section{The Image Size and Synchrotron Self Absorption}
\label{self_abs}

In GRB afterglows, relativistic electrons are accelerated in the advancing
shock wave to a power law distribution of energies,
$dN/d\gamma_e\propto\gamma_e^{-p}$ for $\gamma_e\geq\gamma_m$. For $p>2$,
the minimal Lorentz factor of the electrons is given by
\begin{equation}\label{gamma_m}
\gamma_m=\left(\frac{p-2}{p-1}\right)\frac{m_p}{m_e}\,
\epsilon_e(\Gamma-1)\ ,
\end{equation}
where $\epsilon_e$ is the fraction of the internal energy behind
the shock in relativistic electrons, and $\Gamma$ is the bulk
Lorentz factor of the shocked fluid. There is a spectral break at
$\nu_m=\nu_{\rm syn}(\gamma_m)$, the synchrotron frequency of
electrons with $\gamma_e=\gamma_m$. Another break in the spectrum
occurs at $\nu_c=\nu_{\rm syn}(\gamma_c)$, the synchrotron
frequency of an electron that cools on the dynamical time.

At sufficiently low frequencies, below the self absorption
frequency $\nu_{sa}$, the optical depth to synchrotron self
absorption $\tau_\nu$ becomes larger than unity, causing an
additional break in the spectrum. In this spectral range, the
emitted intensity is given by the Rayleigh-Jeans part of a black
body spectrum, where the black body temperature is taken as the
effective temperature $T_{\rm eff}$ of the electrons that are
emitting the radiation at the observed frequency $\nu$. In the
local rest frame of the emitting fluid this may be written as
\begin{equation}\label{Inu_local}
I'_{\nu'}=\frac{2(\nu')^2}{c^2}k_BT_{\rm
eff}=\frac{2(\nu')^2}{c^2}\gamma_{\rm eff}m_e c^2\ ,
\end{equation}
where primed quantities are measured in the local rest frame of
the emitting fluid while un-primed quantities are measured in the
observer frame (the rest frame of the central source). When
$\nu_{sa}>\nu_m$, the emission at $\nu_m<\nu<\nu_{sa}$ is
dominated by electrons for which $\nu\sim\nu_{\rm
syn}(\gamma_e)\propto\gamma_e^2$, giving $\gamma_{\rm
eff}\propto\nu^{1/2}$ and $F_\nu\propto I_\nu\propto\nu^{5/2}$.
For $\nu_m>\nu_c$ there is rapid cooling and all the electrons cool
significantly within a dynamical time \citep{SPN98}. When
$\nu_m>\max(\nu_c,\nu_{sa})$, then as $\nu$ decreases below
$\nu_{sa}$ the distance $l$ behind the shock where $\tau_\nu(l)=1$
decreases. The electrons in that location, which are responsible
for most of the observed emission, have had less time to cool
after passing the shock and therefore have a higher $T_{\rm
eff}=\gamma_{\rm eff}(m_ec^2/k_B)$. In this case $\gamma_{\rm
eff}\propto 1/l\propto\nu^{-5/8}$ and $F_\nu\propto\nu^{11/8}$
\citep{GPS00}. At a sufficiently small distance behind the shock,
smaller than $l_c$, an electron with an initial Lorentz factor
$\gamma_m$ does not have enough time to cool significantly after
crossing the shock. Therefore, most electrons within a distance of
$l_c$ from the shock have $\gamma_e\sim\gamma_m$, and the
effective temperature in this region is $T_{\rm
eff}\approx\gamma_m m_e c^2/k_B$. At sufficiently low frequencies
\citep[below $\nu_{ac}$, see][]{GPS00} $l$ becomes smaller than
$l_c$ and $\gamma_{\rm eff}\approx\gamma_m$ independent of
$\nu$, and therefore $F_\nu\propto\nu^2$ at $\nu<\nu_{ac}$. For
slow cooling ($\nu_m<\nu_c$), $\gamma_{\rm eff}\approx\gamma_m$
and $F_\nu\propto\nu^2$ immediately below $\nu_{sa}$.

The observed specific intensity is given by $I_\nu=(\nu/\nu')^3
I'_{\nu'}$ and
$\nu'/\nu=(1+z)\Gamma(1-\beta\cos\theta)\sim(1+z)/\Gamma$ where
$z$ is the source redshift and $\theta$ is the angle between the
direction to the observer and the velocity vector of the emitting
material in the observer frame. The observed flux density is
$F_\nu=\int d\Omega\cos\tilde{\theta}I_\nu\approx\Omega I_\nu$ where
$\Omega\approx\pi(R_\perp/D_A)^2=(1+z)^2\pi(R_\perp/D_p)^2=
(1+z)^4\pi(R_\perp/D_L)^2$ and
$\tilde{\theta}\cong\tan\tilde{\theta}=R_\perp/D_A\ll 1$ are the
solid angle and angular radius of the source image, respectively.
Here $R_\perp$ is the radius of the observed image (its apparent
size on the plane of the sky) and $D_A$, $D_p$ and $D_L$ are the
angular, proper and luminosity distances to the source,
respectively. Thus one obtains
$I_\nu\approx[\Gamma/(1+z)]^3[2(\nu')^2/c^2]kT_{\rm eff}\approx
[\Gamma/(1+z)]2\nu^2\gamma_{\rm eff}m_e$, and \citep{KP97}
\begin{equation}\label{Fnu}
F_\nu\approx 2\pi\nu^2 m_e\Gamma\gamma_{\rm
eff}(1+z)\left(\frac{R_\perp}{D_p}\right)^2\ .
\end{equation}

In deriving Eq. (\ref{Fnu}) the specific intensity $I_\nu$ was assumed to
be uniform across the observed image. A more accurate calculation would
have to integrate over the contribution to the observed emission from
different radii $R$ and angles $\theta$ from the line-of-sight for a fixed
observed time $t$ \citep[e.g.,][]{GPS99b}, which results in a non-uniform
$I_\nu$ across the image. Therefore, when using Eq. (\ref{Fnu}) one must
choose some effective value for $I_\nu$ which should correspond to its
average value across the image. Since $I_\nu$ depends on $\Gamma$, this is
equivalent to choosing an effective value of $\Gamma$. Since $\Gamma$
depends on $R$, one also has to find at which $R$ or $\theta$ should the
value of $\Gamma$ be evaluated in Eqs.  (\ref{gamma_m}) and
(\ref{Fnu}). Usually $\nu_{sa}<\nu_m<\nu_c$ in which case $\gamma_{\rm
eff}\approx\gamma_m$ so that $I_\nu$ depends on $\Gamma$ not only through
the Lorentz transformations, but also through the value of $\gamma_m$,
i.e. $\Gamma$ enters into both Eqs. (\ref{gamma_m}) and
(\ref{Fnu}). Comparing Eq.  (\ref{Fnu}) with the more accurate expression
calculated by \citet{GS02} using the Blandford-McKee (1976) self similar
spherical solution, we find that the two expressions are in relatively good
agreement\footnote{We find that the ratio of the numerical coefficient in
Eq. (\ref{Fnu}) to that in \citet{GS02} is in this case $1.09(3p-1)/(3p+2)$
for $k=2$ and $1.71(3p-1)/(3p+2)$ for $k=0$.} if $\Gamma$ is evaluated just
behind the shock at the location where $R_\perp$ is located.  This should
be a good approximation before the jet break time in the light curve,
\begin{eqnarray}\label{t_j}
t_j &=&\frac{(1+z)}{4c}\left[\frac{(3-k)E}{2\pi A
c^2}\right]^{1/(3-k)}\theta_0^2
\\ \nonumber
&\approx &\left\{\begin{matrix}
0.66(1+z)(E_{51}/n_0)^{1/3}(\theta_0/0.1)^2\;{\rm days} & \ \
(k=0) \cr & \cr 0.34(1+z)(E_{51}/A_*)(\theta_0/0.1)^2\;{\rm days}
& \ \ (k=2)\end{matrix}\right.\ .
\end{eqnarray}
At $t>t_j$, however, it is less clear how well this approximation holds,
and it might be necessary to evaluate $\Gamma$ at a different location. In
particular, as we shall see below, this approximation does not work well
for {\bf model 2} with $k=0$ where $\Gamma$ need to be evaluated
near the head of the jet, rather than at the side of the jet where
$R_\perp$ is located.

\begin{figure}
\plotone{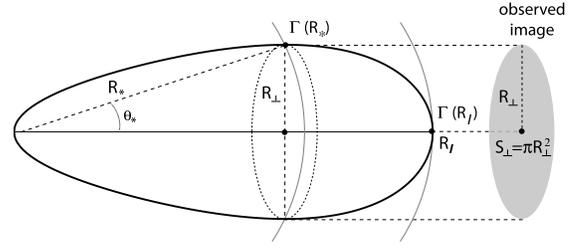}
\caption{\label{image_size} Schematic illustration of the equal-arrival
time surface (thick black line), namely the surface from where the photons
emitted by the shock front arrive at the same time to the observer (far on
the right-hand-side). The maximal lateral extent of the observed image,
$R_\perp$, is located at an angle $\theta_*$, where the shock radius and
Lorentz factor are $R_*$ and $\Gamma_*=\Gamma_{\rm sh}(R_*)$,
respectively. The area of the image on the plane of the sky is $S_\perp=\pi
R_\perp^2$. The shock Lorentz factor $\Gamma_{\rm sh}$ varies with $R$ and
$\theta$ along the equal-arrival time surface. The maximal radius, $R_l$,
on the equal-arrival time surface is located along the line-of-sight. If,
as expected, $\Gamma_{\rm sh}$ decreases with $R$, then
$\Gamma_l=\Gamma_{\rm sh}(R_l)$ is the minimal shock Lorentz factor on the
equal-arrival time surface.}
\end{figure}

The image size is given by $R_\perp=\max(R\sin\theta)$ along the
equal arrival time surface (see Figure \ref{image_size}). The
equal arrival time surface is the surface from where photons that
are emitted at the shock front arrive to the the observer
simultaneously. Since the emission originates only from behind the
shock front, the projection of the equal arrival time surface onto
the plane of the sky (i.e. the plane perpendicular to the
line-of-sight) determines the boundaries of the observed image,
and its apparent size (see Figure \ref{image_size}). For a
spherical shock front with any $R(t_{\rm lab})$,
$R_\perp=\max(R\sin\theta)$ is located at an angle $\theta_*$
which satisfies $\cos\theta_*=\beta_*$ (see Appendix
\ref{theta_*}), where $\beta_*$ and
$\Gamma_*=(1-\beta_*^2)^{-1/2}$ are the velocity (in units of $c$)
and the Lorentz factor of the shock front\footnote{Note that we
use $\beta_{\rm sh}$ or $\Gamma_{\rm sh}$ for the location of the
emitting fluid, which is always just behind the shock. On the
other hand, we use $\beta$ or $\Gamma$ (which are slightly smaller
than $\beta_{\rm sh}$ or $\Gamma_{\rm sh}$, respectively) for the
Lorentz transformations of the emitted radiation, since these are
the bulk velocity and Lorentz factor of the emitting fluid.} at
$\theta_*$. This implies that $R_\perp(t)= R_*(t)/\Gamma_*(t)$
where $R_*(t)=R(t,\cos\theta=\beta_*)$ is the radius of the shock
at $\theta_*=\arccos\beta_*$. Therefore $\Gamma_*=\Gamma_{\rm
sh}(R_*)$ and $\beta_*=\beta_{\rm sh}(R_*)$.  Although the shock
front is probably not simply a section of a sphere
\citep{Granot01}, we consider this as a reasonable approximation
for our purpose. The expression for $\theta_*$ in the more general
case of an axially symmetric shock is given in Appendix
\ref{theta_*}.

The apparent speed, $\beta_{\rm ap}=[(1+z)/c](dR_\perp/dt)$,
has a simple form for a point source moving at an angle $\theta$
from our line-of-sight, $\beta_{\rm ap}=\beta_{\rm
sh}\sin\theta/(1-\beta_{\rm sh}\cos\theta)$. Substituting
$\cos\theta=\beta_{\rm sh}$ in this expression gives $\beta_{\rm
ap}=\Gamma_*\beta_*=\sqrt{\Gamma_*^2-1}$ or
$\Gamma_*=\sqrt{1+\beta_{\rm ap}^2}$ and $\beta_*=\beta_{\rm
ap}/\sqrt{1+\beta_{\rm ap}^2}$. In Appendix \ref{app_vel} we show
that this result holds for any spherically symmetric shock front,
and we also generalize it to an axially symmetric shock. Finally, the
Lorentz factor $\Gamma$ of the shocked fluid just behind the shock
at $\theta_*$ is related to the Lorentz factor of the shock
itself, $\Gamma_*$, by
$\Gamma_*^2=(\Gamma+1)[\hat{\gamma}(\Gamma-1)+1]^2/
[\hat{\gamma}(2-\hat{\gamma})(\Gamma-1)+2]$ \citep{BM76} where
$\hat{\gamma}$ is the adiabatic index of the shocked fluid. For
$\Gamma_*\gg 1$, $\hat{\gamma}=4/3$ and
$\Gamma=\Gamma_*/\sqrt{2}$.

\section{The Temporal Evolution of the Image Size}
\label{temp_ev}

For simplicity, we consider a uniform GRB jet with sharp edges and a
half-opening angle $\theta_j$, with an initial value of $\theta_0$. The
evolution of the angular size of the image and its angular displacement
from the central source on the plane of the sky, for viewing angles
$\theta_{\rm obs}>\theta_0$ from the jet axis, was outlined in
\citet{GL03}. Here we expand this discussion to include viewing angles
within the initial jet opening angle, $\theta_{\rm obs}<\theta_0$, for
which there is a detectable prompt gamma-ray emission (similarly to GRB
030329 which is considered in the next section). For $\theta_{\rm
obs}<\theta_0$, $R_\perp$ is the observed size of the image, while for
$\theta_{\rm obs}>\theta_0$ it represents the displacement with respect to
the central source on the plane of the sky.

\begin{figure}
\plotone{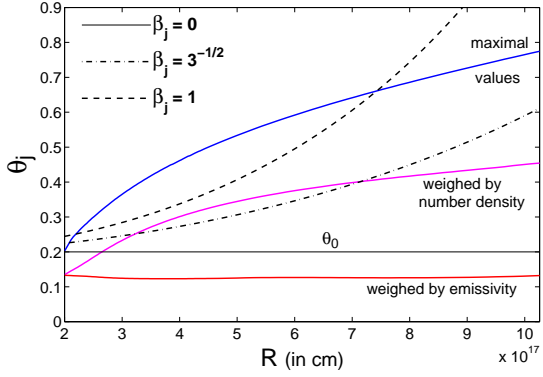} \caption{\label{lateral_exp} Evolution
of the jet half-opening angle $\theta_j$ as a function of radius
$R$, for various illustrative cases. The solid line shows the
evolution derived from 2D hydrodynamical simulations
\citep{Granot01}. The different lines give the maximal polar angle
$\theta$ of the shock front (which is obtained at a relatively
small radius where a minor fraction of the emission is produced),
and the average values of $\theta$ within the jet when averaged
over the circumburst gas density and over the total emissivity.
Most of the emission comes from within the original jet opening
angle, $\theta_0=0.2$.  Also shown is the evolution of
$\theta_j(R)$ predicted by simple semi-analytic models.  Three
illustrative cases are depicted where the lateral expansion
speed is assumed to be $\beta_j=0$, $3^{-1/2}$ and $1$ in the
local rest frame \citep{Rhoads99,SPH99,ONP04}. Since the onset of
lateral expansion in the simple models takes place at a somewhat
larger radius ($R_j$), a higher value of gas density is adopted
for these models in order to show more easily their different
qualitative behaviors.}
\end{figure}

In this section we concentrate on a viewing angle along the jet
axis, $\theta_{\rm obs}=0$, and in the next section we briefly
outline the expected differences for $0<\theta_{\rm
obs}<\theta_0$. For $\theta_{\rm obs}=0$, the observed image is
symmetric around the line-of-sight (to the extent that the jet is
axisymmetric). At $t<t_j$ the edge of the jet is not visible and
the observed image is the same as for a spherical flow:
$R_\perp\propto (E_{\rm iso}/A)^{1/2(4-k)}t^{(5-k)/2(4-k)}\propto
(E/A)^{1/2(3-k)}t_j^{-1/2(4-k)}t^{(5-k)/2(4-k)}$ for an external
density profile $\rho_{\rm ext}=Ar^{-k}$, i.e. $a=(5-k)/2(4-k)$
where $a\equiv d\ln R_\perp/d\ln t$. Here $E$ is the true kinetic
energy of the jet, and $E_{\rm iso}=f_b^{-1}E$ is the isotropic
equivalent energy where $f_b=1-\cos\theta_0\approx\theta_0^2/2$
is the beaming factor. At $t<t_j$ the flow is described by
the Blandford-McKee (1976) self similar solution, which provides
an accurate expression for the image size \citep{GPS99a,GS02},
\begin{eqnarray}\nonumber
R_\perp &=& \left[\frac{2^{2-k}(17-4k)(4-k)^{5-k}E_{\rm
iso}c^{3-k}t^{5-k}}{\pi(5-k)^{5-k}(1+z)^{5-k}A}\right]^{1/2(4-k)}
\\ \nonumber
\\ \label{R_perp_sph}
&=& \left\{\begin{matrix} 3.91\times 10^{16}(E_{\rm
iso,52}/n_0)^{1/8}[t_{\rm days}/(1+z)]^{5/8}\;{\rm cm} & \ \ (k=0)
\cr & \cr 2.39\times 10^{16}(E_{\rm iso,52}/A_*)^{1/4}[t_{\rm
days}/(1+z)]^{3/4}\;{\rm cm} & \ \ (k=2)\end{matrix}\right.\ .
\end{eqnarray}
At $t>t_{\rm NR}$ the jet gradually approaches the
Sedov-Taylor self similar solution, asymptotically reaching
$R_\perp\propto (Et^2/A)^{1/(5-k)}$, i.e. $a=2/(5-k)$. At
$t_j<t<t_{\rm NR}$ there is a large uncertainty in the
hydrodynamical evolution of the jet, and in particular its rate
of sideways expansion. We therefore consider two extreme
assumptions which should roughly bracket the different possible
evolutions of $R_\perp(t)$: {\bf (1)} relativistic
lateral expansion in the comoving frame \citep{Rhoads99,SPH99},
for which $\theta_j\approx\max(\theta_0,\gamma^{-1})$ so that at
$t_j<t<t_{\rm NR}$ we have
$\gamma\approx\theta_j^{-1}\approx\theta_0^{-1}\exp(-R/R_j)$, and
{\bf (2)} little or no lateral expansion,
$\theta_j\approx\theta_0$ for $t<t_{\rm NR}$, in which case
appreciable lateral expansion occurs only when the jet becomes
sub-relativistic and gradually approaches spherical symmetry. We
shall refer to these models as {\bf model 1} and {\bf model 2},
respectively. {\bf Model 2} is also motivated by the results of
numerical simulations (see Figure \ref{lateral_exp}) which show
only modest lateral expansion as long as the jet is relativistic
\citep{Granot01,KG03,CGV04}. These numerical results are also
supported by a simple analytic argument that relies on the shock
jump conditions for oblique relativistic shocks \citep{KG03}.

\begin{figure}
\plotone{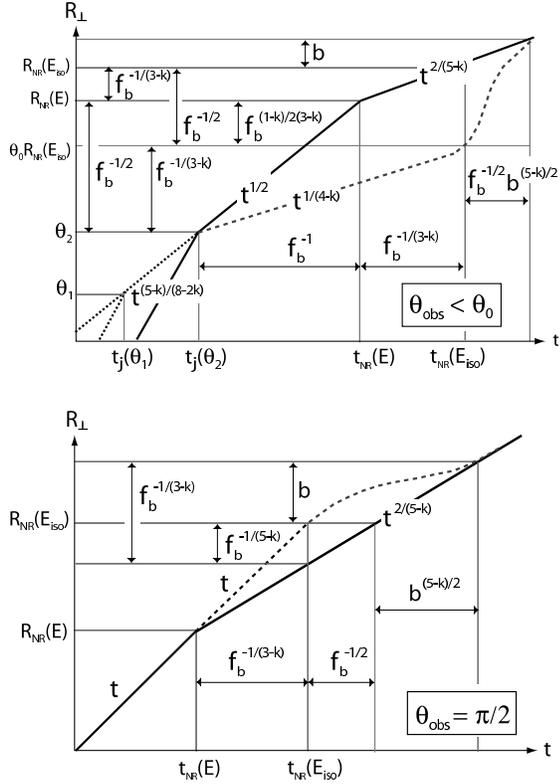}
\caption{\label{R_perp_t} Schematic plot of
the evolution of the observed afterglow image size $R_\perp$ of a
uniform GRB jet with sharp edges. The jet is either viewed from
within the initial jet opening angle, $\theta_{\rm obs}<\theta_0$
({\it upper panel}), or from $\theta_{\rm obs}\approx 90^\circ$
({\it lower panel}). The solid line is for {\bf model 1}
(relativistic lateral expansion in the local rest frame) and the
dashed line is for {\bf model 2} (little or no lateral expansion
before $t_{\rm NR}$).  The dotted line in the upper panel
represents jets (in {\bf model 2}) with a smaller $\theta_0$ and
the same true energy $E$, which converge to the same self similar
dynamics and therefore the same $R_\perp(t)$ after the jet break
time $t_j$.  Also shown are the ratios of various values of
$R_\perp$ and $t$.}
\end{figure}

Figure \ref{R_perp_t} schematically shows the evolution of $R_\perp(t)$ for
these two extreme models, both when viewed on-axis ($\theta_{\rm
obs}<\theta_0$) as required for seeing the prompt gamma-ray emission, and
for $\theta_{\rm obs}\approx 90^\circ$ as will typically be the case for
GRB jets found in nearby SNe Type Ib/c \citep{Pa01,GL03,GR-R04,RM04}. For
$\theta_{\rm obs}<\theta_0$ at $t_j<t<t_{\rm NR}$ we have
$R_\perp\propto(E/A)^{1/2(3-k)}t^{1/2}$ for {\bf model 1}, and
$R_\perp\propto(E_{\rm
iso}t/A)^{1/(4-k)}\propto(E/A)^{1/(3-k)}(t/t_j)^{1/(4-k)}$ for {\bf model
2}. Therefore, with $k=2$ we have $a=1/2$ for both models, despite their
very different jet dynamics. For $k=0$ we have $a=1/2$ for {\bf model 1}
and $a=1/4$ for {\bf model 2}.

In {\bf model 1}, the jet is already relatively close to being
spherical (i.e. $\theta_j\sim 1$) at $t_{\rm NR}=t_{\rm NR}(E)$,
where $R_{\rm NR}(E)=ct_{\rm NR}(E)=[(3-k)E/4\pi Ac^2]^{1/(3-k)}$,
and its radius is similar to that of the Sedov-Taylor solution ,
$R_{\rm ST}(E,t)=\xi(Et^2/A)^{1/(5-k)}$, corresponding to the same
time $t$, where $\xi=\xi(k,\hat{\gamma})\sim 1$. Therefore, we
expect it to approach spherical symmetry on a few dynamical times,
i.e. when the radius increases by a factor of $b\sim{\rm a\ few}$,
corresponding to a factor of $\sim b^{(5-k)/2}$ in time, and the
transition between the asymptotic power laws in $R_\perp(t)$ is
expected to be smooth and monotonic.

In {\bf model 2}, however, the jet becomes sub-relativistic only
at $R_{\rm NR}(E_{\rm iso})=ct_{\rm NR}(E_{\rm iso})$, which is a
factor of $\sim(E_{\rm iso}/E)^{1/(3-k)}=f_b^{-1/(3-k)}\sim
\theta_0^{-2/(3-k)}$ larger than $R_{\rm NR}(E)=ct_{\rm NR}(E)$
and a factor of $\sim f_b^{-1/(5-k)}\sim\theta_0^{-2/(5-k)}$
larger than $R_{\rm ST}[E,t_{\rm NR}(E_{\rm iso})]$. It also keeps
its original opening angle, $\theta_j\approx\theta_0$ until
$t_{\rm NR}(E_{\rm iso})$, and hence at this time the jet is still
very far from being spherical. Thus, once the jet becomes
sub-relativistic, we expect it to expand sideways significantly,
and become roughly spherical only when it has increased its radius
by a factor of $b\sim{\rm a\ few}$. This should occur, however,
roughly at a time $t_{\rm sph}$ when $R_{\rm ST}(E,t_{\rm
sph})=bR_{\rm NR}(E_{\rm iso})$, i.e.
\begin{equation}\label{t_sph}
t_{\rm sph}/t_{\rm NR}(E_{\rm iso})\approx
f_b^{-1/2}b^{(5-k)/2}\approx\sqrt{2}\,\theta_0^{-1}b^{(5-k)/2}\ .
\end{equation}
This is a factor of $\sim f_b^{-1/2}\approx 14(\theta_0/0.1)^{-1}$
larger than the expected transition time in {\bf model 1}, and for
$b\sim 2-3$ gives a factor of $\sim(80-220)(\theta_0/0.1)^{-1}$
for $k=0$ and $\sim (40-70)(\theta_0/0.1)^{-1}$ for $k=2$. During
this transition time, $R_\perp(\theta_{\rm obs}<\theta_0)$ grows
by a factor of $\sim f_b^{-1/2}b\sim\theta_0^{-1}b$ while
$R_\perp(\theta_{\rm obs}\approx 90^\circ)$ grows by a factor of
$\sim b$. This implies that during the transition,
\begin{equation}\label{a_av}
\langle a\rangle = \left\{\begin{matrix}\frac{\ln b-(1/2)\ln
f_b}{[(5-k)/2]\ln b-(1/2)\ln f_b} & (\theta_{\rm obs}<\theta_0)
\cr & \cr \frac{\ln b}{[(5-k)/2]\ln b-(1/2)\ln f_b} & (\theta_{\rm
obs}=90^\circ)\end{matrix}\right.\ ,
\end{equation}
and $0<\langle a\rangle<2/(5-k)$ for $\theta_{\rm obs}<\theta_0$
while $2/(5-k)<\langle a\rangle<1$ for $\theta_{\rm
obs}=90^\circ$, where $\langle a\rangle\approx 2/(5-k)$ in the
limit $b\gg\theta_0^{-1}$ (which is not very realistic). The other
limiting value of $\langle a\rangle\approx 0$ for $\theta_{\rm
obs}<\theta_0$ and $\langle a\rangle\approx 1$ for $\theta_{\rm
obs}=90^\circ$ is approached in the limit $b\ll\theta_0^{-1}$.
Typical parameter values ($b\sim 2-3$, $\theta_0\sim 0.05-0.2$)
are somewhat closer to the latter limit. For example, for $k=0$,
$b=2.5$ and $\theta_0=0.1$ we have $\langle a\rangle\approx 0.722$
for $\theta_{\rm obs}<\theta_0$ and $\langle a\rangle\approx
0.185$ for $\theta_{\rm obs}=90^\circ$. This demonstrates that for
on-axis observers there should be a sharp rise in $R_\perp$, while
for observers at $\theta_{\rm obs}\approx 90^\circ$ there should
be a very moderate rise in $R_\perp$ during the transition phase
from the asymptotic $t_j\ll t\ll t_{\rm NR}$ and $t\gg t_{\rm NR}$
regimes. Furthermore, as is illustrated in Figure \ref{R_perp_t},
this transition would not be monotonic in {\bf model 2}. This is
because during the transition $a$ passes through values larger
(smaller) than both of its asymptotic values for $\theta_{\rm
obs}<\theta_0$ ($\theta_{\rm obs}\approx 90^\circ$).

For comparison, and in order to perform a quantitative comparison with the
data, we consider a simple semi-analytic model where the shock front at any
given lab frame time occupies a section of a sphere within
$\theta<\theta_j$, and $R_\perp$ is located at
$\theta_\perp=\min(\theta_*,\theta_j)$. The observer time assigned to a
given $\theta_\perp(t_{\rm lab})$ is $t=t_{\rm lab}-[R(t_{\rm
lab})/c]\cos\theta_\perp(t_{\rm lab})$. We follow \citet{ONP04} with minor
differences: {\it (i)} we choose the normalization of $R_\perp$ at $t\ll
t_j$ so that it will coincide with the value given by the Blanford-McKee
solution (i.e. Eq. \ref{R_perp_sph}), and {\it (ii)} the lateral spreading
verlocity in the comoving frame, $\beta_j$, for model 2 smoothly varies
from $\beta_j\ll 1$ at $t\ll t_{\rm NR}$ to the sound speed,
$\beta_j\approx c_s/c$, at $t>t_{\rm NR}$. The latter is achieved by taking
$\beta_j$ to be the sound speed suppressed by some power of $\Gamma$.

\begin{figure}
\plotone{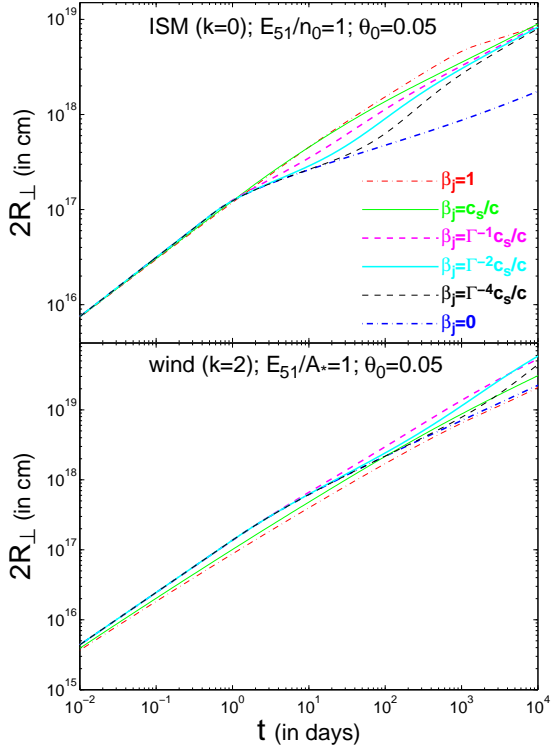} \caption{\label{models12} Evolution of the
source size (or more precicely, its diameter $2R_\perp$) as a
function of time, for a uniform density environment ($k=0$, {\it
upper panel}) and for a stellar wind ($k=2$, {\it lower panel}).
Different recipes are considered for the lateral spreading
velocity in the comoving frame, $\beta_j$. See text for more
details.}
\end{figure}

Figure \ref{models12} shows the resulting $R_\perp(t)$ for ISM
($k=0$) and stellar wind ($k=2$) environments, and different
recipes for $\beta_j$. For a given $\beta_j$ recipe,
$R_\perp(t)$ depends on $E/A$ and $\theta_0$. The values of these
parameters that were used in Figure \ref{models12} are indicated
in the figure. For $k=2$ the spread in $R_\perp(t)$ for the
different $\beta_j$ recipes is smaller than for $k=0$. This is
understandable since the asymptotic values of $a$ are the
same for models 1 and 2. There is still a non-negligible spread,
however, as the asymptotic value of $a=1/2$ at $t_j\ll t\ll t_{\rm
NR}$ is not reached.\footnote{This is since it takes a long time
to approach this limit for $k=2$, which is longer than the
dynamical range between $t_j$ and $t_{\rm NR}$.} At $t\gg t_{\rm
NR}$ all recipes for $\beta_j$ approach the same value of
$R_\perp(t)$, except for $\beta_j=0$ for which $R_\perp(t)$ is
smaller by a factor of $\sin\theta_0$. For $\beta_j=0$ and $k=0$
there is a pronounced flattening in $R_\perp(t)$ at $\sim
1.2\;$day, which is a factor of $\sim 7$ larger than the value of
$t_j=0.165\;$days that is implied by Eq. (\ref{t_j}). We must stress
that this simple model becomes unrealistic around $t_{\rm
NR}$.

The apparent velocity of a point source is $\beta_{\rm
ap}=\beta\sin\theta/(1-\beta\cos\theta)$. For $\theta_{\rm obs}>\theta_0$,
as long as $\theta_j<\theta_{\rm obs}$ and $t<t_{\rm NR}$ we have
$\beta_{\rm ap}\approx 2\Gamma_{\rm sh}^2\theta/[1+(\Gamma_{\rm
sh}\theta)^2]\approx 2/\theta$. For $\theta_{\rm obs}=\pi/2$ we have
$\beta_{\rm ap}=\beta_{\rm sh}$ which is close to 1 at $t<t_{\rm NR}$. For
$\Gamma_{\rm sh}\gg 1$ and $\theta>\Gamma_{\rm sh}^{-1}$ we have
$\beta_{\rm ap}\approx\sin\theta/(1-\cos\theta)$, so that $\beta_{\rm
ap}>1$ for $\theta_{\rm obs}<\pi/2$ and $\beta_{\rm ap}<1$ for $\theta_{\rm
obs}>\pi/2$ \citep[i.e. for the counter jet, assuming a double sided jet;
see Figure 2 of][]{GL03}. For $\theta_{\rm obs}<\theta_0$ we have
$\beta_{\rm ap}=\Gamma_*\beta_*\approx\Gamma_*$ at $t<t_{\rm NR}$. At
$t<t_j$ we get $\theta_\perp=\theta_*<\theta_0$ and the shock front is
roughly spherical with an approximately uniform Lorentz factor within
$\theta\lesssim\theta_*$, so that $\Gamma_*\approx\Gamma_{\rm sh}$. At
$t_j<t<t_{\rm NR}$ we have $\theta_*\approx\theta_j\approx\Gamma_{\rm
sh}^{-1}$ and $\beta_{\rm ap}\approx\Gamma_*\approx\Gamma_{\rm sh}$ for
{\bf model 1}, suggesting that using $\Gamma(\theta_*)$ for calculating the
emission (i.e. in Eqs. \ref{gamma_m}, \ref{Fnu} and \ref{eps_e}) is a
reasonable approximation. For {\bf model 2},
$\theta_*\approx\theta_j\approx\theta_0$ and $\beta_{\rm
ap}\approx\Gamma_*\approx 2\theta_0\Gamma_{\rm sh}^2$, so
that\footnote{Here $\Gamma_{\rm sh}$ represents the uniform shock Lorentz
factor in the simple semi-analytic model described at the end of \S
\ref{self_abs}, where the shock at any given $t_{\rm lab}$ occupies a
section of a sphere and abruptly ends at $\theta_j$, and at $t_j<t<t_{\rm
NR}$ $R_\perp$ is located at $\theta_j$. On the other hand,
$\Gamma_*=\Gamma_{\rm sh}(\theta_*)$ is the Lorentz factor at the angle
$\theta_*$ where $R_\perp$ is located for a smooth and continuous (and
therefore more realistic) shock front, for which $\Gamma_{\rm sh}$ changes
with $\theta$ at a given $t_{\rm lab}$.} $\Gamma_*/\Gamma_{\rm sh}\approx
2\theta_0\Gamma_{\rm sh}<1$ suggesting that $\Gamma(\theta_*)$
underestimates the effective value of the emissivity-weighted $\Gamma$ ,
which enters the expressions for the observed emission. This results from
the fact that in {\bf model 2} most of the emission at $t_j<t<t_{\rm NR}$
originates from $\theta<\theta_0$ where $\Gamma$ is higher than at
$\theta_*\gtrsim\theta_0$ \citep[see Figure \ref{lateral_exp}
and][]{Granot01}.

\section{Linear Polarization}
\label{pol}

For $0<\theta_{\rm obs}<\theta_0$ the image would not be symmetric around
the line-of-sight, but its typical angular size would be similar to that of
$\theta_{\rm obs}=0$. If there is significant lateral spreading at $t>t_j$,
then this should cause the image to become more symmetric around our
line-of-sight with time. This, by itself, might be a possible diagnostic
for the degree of lateral spreading. The degree of asymmetry in the
observed image should also be reflected in the degree of linear
polarization, and its temporal evolution. While the image might be resolved
only for a very small number of sufficiently nearby GRBs, the linear
polarization might be measured for a larger fraction of GRBs.

Contrary to naive expectations, for very slow lateral expansion
($\beta_j\ll 1$) the polarization decays faster after its peak at $t\sim
t_j$ compared to lateral expansion at the local sound speed,
$\beta_j=c_s/c\approx 3^{-1/2}$, in the comoving frame \citep{Rossi04}. A
very fast lateral expansion in the local frame close to the speed of light
($\beta_j\approx 1$), leads to $\theta_j\approx\max(\theta_0,\gamma^{-1})$
and to three peaks in the polarization light curve, where the polarization
position angle changes by $90^\circ$ as the degree of polarization passes
through zero between the peaks \citep{Sari99}. When there is a slower
lateral expansion or no lateral expansion at all \citep{GL99}, there are
only two peaks in the polarization lightcurve where again the polarization
position angle changes by $90^\circ$ as the degree of polarization passes
through zero between the peaks. The peak polarization is higher for
$\beta_j\approx 0$ ($\sim 15\%-16\%$) compared to $\beta_j=3^{-1/2}$ ($\sim
9\%$) \citep{Rossi04}. The maximal observed degree of polarization is,
however, usually $\lesssim 3\%$ suggesting that the magnetic field
configuration behind the shock is more isotropic than a random field fully
within the plane of the shock \citep{GK03} which is expected if the
magnetic field is produced by the Weibel instability \citep{ML99}. This
changes the overall normalization of the polarization light curve, and
hardly affects its shape \citep{GK03}. Since the overall normalization is
the most pronounced difference between slow and fast lateral expansion, and
it is very similar to the effect of the degree of anisotropy of the
magnetic field behind the shock, it would be very hard to constrain the
degree of lateral expansion from the polarization light curves. There are
also other possible complications, such as a small ordered magnetic field
component \citep{GK03} which can induce polarization that is not related to
the jet structure.

\citet{Taylor04} put a $3\;\sigma$ upper limit of $1\%$ on the
linear polarization in the radio ($\nu=8.4\;$GHz) at
$t=7.71\;$days. They attribute the low polarization to synchrotron
self absorption. Indeed, $\nu_{sa}$ is above $8.4\;$GHz at this
time, but only by a factor of $\sim 2$. One might expect a
suppression of the polarization in the self absorbed region of the
synchrotron spectrum since it should follow the Rayleigh-Jeans
part of a black body spectrum, and depend only on the electron
distribution (i.e. the ``effective temperature") and not on the
details of the magnetic field \citep{GPS99b}. The optical depth to
self absorption does, however, depend on the details of the
magnetic field, and may thereby vary with the direction of
polarization. Therefore, there might still be polarization at
$\nu\lesssim\nu_{sa}$ which will go to zero in the limit
$\nu\ll\nu_{sa}$. An ordered magnetic field in the
shocked fluid through which the emitted synchrotron radiation
propagates on its way to the observer, might induce some
polarization in the observed radiation \citep{SWL04}. These effects
are suppressed roughly by a factor of the square root of the ratio
between the magnetic field coherence length and the width of the
emitting region (which is of the order of the typical path
length of an emitted photon through the shocked plasma before it
escapes the system).

\section{The Surface Brightness Profile}
\label{SB}

\citet{Taylor04} use a circular Gaussian profile for their quoted
values, and also tried a uniform disk and thin ring. They find
that a Gaussian with an angular diameter size of $1\;$mas is
equivalent to a uniform disk with an angular diameter of
$1.6\;$mas and a thin ring with an angular diameter of $1.1\;$mas.
At $t<t_j$ the jet dynamics are close to that of a spherical flow,
since the center of the jet is not in causal contact with its
edge, and the dynamics can be described by the Blandford-Mckee
(1976) spherical self similar solution (within the jet, at
$\theta<\theta_0$). The surface brightness in this case has been
investigated at length in several works
\citep{Waxman97,Sari98,PM98,GPS99a,GPS99b,GrL01}. The surface
brightness profile of the image, normalized by its average
value across the image, is the same  within each power law segment
of the spectrum, but changes between different power law segments
\citep{GrL01}. The afterglow image is limb brightened, resembling
a ring, in the optically thin part of the spectrum and more
uniform, resembling a disk, at the self absorbed part of the
spectrum. This can affect the angular size of the image that is
inferred from the observations \citep{Taylor04}, where the angular
diameter for a uniform disk (thin ring) is a factor of 1.6 (1.1)
larger than the values quoted by \citet{Taylor04} for a circular
Gaussian surface brightness profile. This effect would be more
important at $\nu\lesssim\nu_{sa}$ where the afterglow image
resembles a uniform disk rather than a thin ring.

\begin{figure}
\plotone{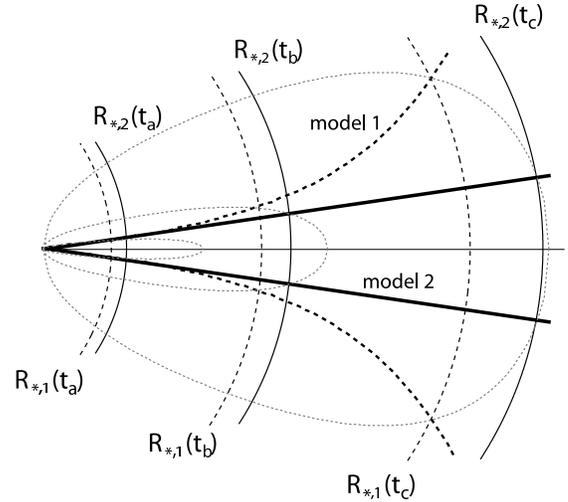}
\caption{\label{EAT_surface} Evolution of
the source size as a function of time for $t>t_j$. In {\bf model
1}, the lateral expansion in the local frame is relativistic while
in {\bf model 2} there is no lateral expansion at $t>t_j$. The
shock front at any given time $t_{\rm lab}$ is assumed to be part
of a sphere which abruptly ends at a finite angle $\theta_j$ from
the jet axis. The gray dotted lines represent the equal arrival
time surfaces at three different observed times. Since the jet
dynamics, $R(t_{\rm lab})$, are different for models {\bf 1} and
{\bf 2}, the equal arrival time surfaces should be different (but
in this sketch, for simplicity, we depicted them as being equal).
At $t>t_j$ (where $\theta_j(R_*)<\theta_*$), the edge of the image
which determines the image size is located at the edge of the jet,
i.e. at an angle $\theta_j$ instead of $\theta_*$.}
\end{figure}

One should keep in mind that the image size of GRB 030329 was inferred well
after the jet break time, $t\gg t_j$, and relatively close to the
non-relativistic transition time, $t\sim t_{\rm NR}$.  However, at
$t_j<t<t_{\rm NR}$ the jet dynamics is poorly known, and this uncertainty
must necessarily be reflected in any calculation of the afterglow image at
this stage, which could only be as good as the assumed dynamical model for
the jet. The afterglow image at this stage was calculated by \citet{IN01}
assuming lateral expansion at the local sound speed \citep{Rhoads99},
similar to our {\bf model 1}. They find that at $t<t_j$ the surface
brightness diverges at the outer edge of the image, which is an artifact of
their assumption of emission from a two dimensional surface
\citep{Sari98,GrL01} identified with the shock front. Calculating the
contribution from all the volume of the emitting fluid behind the shock
makes this divergence go away, except for certain power law segments of the
spectrum where the emission indeed arises from a very thin layer just
behind the shock \citep{GrL01}.  At $t>t_j$ \citet{IN01} obtain a
relatively uniform surface brightness profile. However, this is due to the
unphysical assumption that the shock front at any given lab frame time is
part of a sphere within some finite angle $\theta_j$ from the jet symmetry
axis where the jet ends abruptly. The edge of the image in this case
corresponds to this un-physical point where the jet ends abruptly (see
Figure \ref{EAT_surface}). More physically, as is shown by numerical
simulations \citep{Granot01}, the shock front is not a section of a sphere
and is instead round without any sharp edges. Similarly to the
spherical-like evolution at $t<t_j$, the edge of the image would in this
case correspond to $R_\perp=\max(R\sin\theta)$, and thus the image is
expected to be limb brightened for the same qualitative reasons that apply
at $t<t_j$, even though there would be some quantitative differences.  A
proper calculation of the afterglow image at $t>t_j$ requires full
numerical simulations of the jet dynamics.

\section{Application to GRB 030329}
\label{application}

We now apply the expressions derived in the previous section to GRB 030329
which occurred at a redshift of $z=0.1685$. We use the image angular
diameter size of $\theta_s\approx 70\;{\rm \mu}$as for\footnote{Throughout
this paper we assume $\Omega_M=0.27$, $\Omega_\Lambda=0.73$ and
$H_0=71\;{\rm km\;s^{-1}\;Mpc^{-1}}$.}  $D_A\approx 587\;$Mpc that was
inferred at $t= 24.5\;$days \citep{Taylor04}, which corresponds to
$R_\perp\approx 0.1\;$pc.  This implies an average apparent speed of
$\langle\beta_{\rm ap}\rangle=(1+z)R_\perp/ct\approx 5.66$. The
instantaneous apparent speed is given by $\beta_{\rm
ap}\equiv[(1+z)/c]dR_\perp/dt=a\langle\beta_{\rm ap}\rangle$ where $a\equiv
d\ln R_\perp/d\ln t$. For GRB 030329, if we also take into account the
inferred source size of $\theta_s\approx 170\;{\rm \mu}$as or
$R_\perp\approx 0.25\;$pc at $t=83.3\;$days and the $2\;\sigma$ upper limit
of $\theta_s<100\;{\rm \mu}$as or $R_\perp<0.14\;$pc at $t=51.3\;$days
\citep{Taylor04}, we have\footnote{Applying the Bayesian inference
formalism developed by \citet{Reich01}, we determine values and
uncertainties for the model parameter $a$. Bayesian inference formalism
deals only with measurements with Gaussian error distributions, not with
lower or upper limits. However, this formalism can be straightforwardly
generalized to deal with limits as well, using two facts: (1) a limit can
be given by the convolution of a Gaussian distribution and a Heaviside
function; and (2) convolution is associative.}  $a=0.71_{-0.3}^{+0.4}$
($1\;\sigma$). This value is between $t=24.5\;$days and $83.3\;$days,
assuming that $R_\perp(t)$ followed a perfect power law behavior $\propto
t^a$ with $a={\rm const}$ during this time. This is a reasonable
approximation for {\bf model 1} or {\bf model 2} with $k=2$ for which
$a=1/2$ at $t_j<t<t_{\rm NR}$ and therefore these models are consistent
with the observed temporal evolution of the image size. For {\bf model 1}
with $k=0$ (see \S \ref{temp_ev}) $a=1/4$ at $t_j\ll t\ll t_{\rm NR}$ but
its value is expected to increase significantly near $t_{\rm NR}$ which we
find to be at $\sim 200\;$days for this model (see Table 1). Therefore, it
can still account for the observed image size at $t=24.5\;$days and
$83.3\;$days together with the upper limits at $51.3\;$days. At
$t=24.5\;$days, however, we still expect the value of $a$ in {\bf model 2}
with $k=0$ to be relatively close to its asymptotic value of $a=1/4$.

\begin{figure}
\plotone{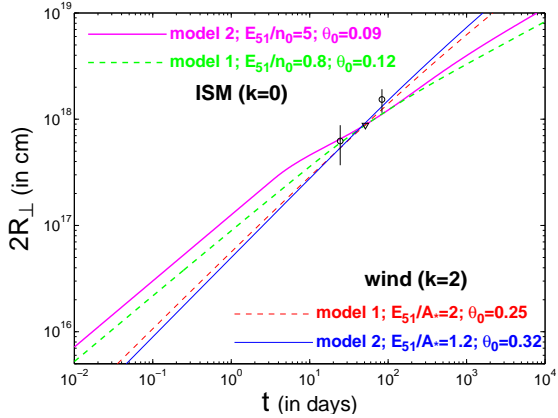} \caption{\label{R_perp_fit} A tentative
fit of a simple semi-analytic realization of {\bf models} 1 and 2
to the observed image size (of diameter $2R_\perp$). The physical
parameters and external density profile for each model are
indicated.}
\end{figure}

Figure \ref{R_perp_fit} shows crude fits between the simple semi-analytic
realization of {\rm models} 1 and 2 (that is described at the end of \S
\ref{temp_ev}) and the observed image size \citep{Taylor04}. For {\bf model
2} we have used the recipe $\theta_j=\Gamma^{-1}(c_s/c)$ for the lateral
expansion. We have treated the value of $E/A$ as a free parameter whose
value was varied in order to get a good fit, while the value of $\theta_0$
was determined according to the observed jet break time $t_j\approx
0.5\;$days using Eq. (\ref{t_j}). In the latter procedure we take into
account an increase in energy by a factor of $f\sim 10$ due to refreshed
shocks \citep{GNP03} between $t_j$ and the times when the image size was
measured. For simplicity, we do not include the effect of the energy
injection on the early image size. The image size that is calculated in
this way to not valid before the end of the energy injection episode (after
several days), but it should be reasonably accurate at $t\gtrsim 25\;$days
when its value had been measured. The values of $E/A$ and $\Gamma(25\;d)$
from these fits are indicated in Figure \ref{R_perp_fit} and in Table 1.

For $a=(0.25,0.5,0.75)$ we obtain $\beta_{\rm ap}\approx
(1.4,2.8,4.2)$, $\Gamma_*\approx (1.7,3.0,4.4)$ and
$\Gamma(\theta_*)\approx (1.5,2.4,3.4)$. The values of
$\Gamma(\theta_*)$ are similar to the value of $\Gamma$ that were
obtained from the fit to the observed image size for {\bf model 1}
and {\bf model 2} with $k=2$, but it is smaller for {\bf model 1}
with $k=0$, as expected (see discussion at end of \S
\ref{temp_ev}).

Using the radio data from \citet{Berger03}, we find that
$F_\nu\approx 10\;$mJy at $t\approx 25\;$days and $\nu=4.86\;$GHz
which according to the spectrum at this time is below $\nu_{sa}$.
\citet{Berger03} also estimated the break frequencies at $t\approx
10\;$days to be $\nu_{sa}\approx 19\;$GHz and $\nu_m\approx
43\;$GHz, which is consistent with $\nu<\nu_{sa}<\nu_m$ at $t=
24.5\;$days. A value of $p=2.25$ was inferred for GRB 030329
\citep{Willingale04}. For the power law segment of the spectrum
where $F_\nu\propto\nu^2$ \citep[labeled `B' in Figure 1
of][]{GS02} we have $\gamma_{\rm eff}\approx\gamma_m$ for which
Eqs. (\ref{gamma_m}) and (\ref{Fnu}) imply
\begin{equation}\label{eps_e}
\epsilon_e\approx\frac{1}{2\pi}\left(\frac{p-1}{p-2}\right)
\frac{(1+z)}{\Gamma(\Gamma-1)}\left(\frac{D_A}{
R_\perp}\right)^2\frac{F_\nu}{m_p\nu^2}\ .
\end{equation}
Using the above values for the flux, $R_\perp$, $\Gamma(\theta_*)$
and $p$ for GRB 030329, Eq. (\ref{eps_e}) gives $\epsilon_e\approx
(0.10,0.023,0.0099)$ for $a=(0.25,0.5,0.75)$. These values of
$\epsilon_e$ are somewhat on the low side compared to the values
inferred from broad band afterglow modelling of other afterglows
\citep[e.g.,][]{PK01b}. In Table 1 we show in addition to these
values of $\epsilon_e$, also the values that are obtained when
evaluating $\Gamma$ from the fit to the image size that is shown
in Figure \ref{R_perp_fit}. The largest difference between these
two estimates of $\epsilon_e$ is for {\bf model 1} with $k=0$, for
which evaluating $\Gamma$ from the fit to the observed source size
probably provides a more accurate estimate.

Since Eq. (\ref{eps_e}) relies on a small number of assumptions, it
is rather robust. However, the value of $\epsilon_e$ in equation
(\ref{eps_e}) is very sensitive to the value of $R_\perp$. This is
because $\epsilon_e\propto 1/R_\perp^2\Gamma(\Gamma-1)$ and for
$\Gamma\gg 1$ we have
$\Gamma\approx\Gamma_*/\sqrt{2}\approx\beta_{\rm
ap}/\sqrt{2}=a\langle\beta_{\rm ap}\rangle/\sqrt{2}\propto
R_\perp$ so that $\epsilon_e\propto R_\perp^{-4}$. For example,
$\theta_s=45\;\mu$as ($R_\perp=0.064\;$pc) at $t=24.5\;$days,
which is still within the measurement errors, would imply
$\epsilon_e=(0.61,0.14,0.060)$ for $a=(0.25,0.5,0.75)$. The latter
values, especially for $a\approx 0.5$, are consistent with
 the value found by \citet{Willingale04} from a
broad band fit to the afterglow data: $\epsilon_e=0.24$ and
$0.18<\epsilon_e<0.31$ at the $90\%$ confidence level, and with
the value of $\epsilon_e\approx 0.16$ found by \citet{Berger03}.

\section{Inferring the Physical Parameters from a Snapshot Spectrum
at $t_j<t<t_{\rm NR}$}
\label{phys_par}

For {\bf model 1}, we obtain expressions for the peak flux and
break frequencies at $t_j<t<t_{\rm NR}$ by using the expressions
for $t<t_j$ from \citet{GS02} in order to estimate their values at
$t_j$, and then using the their temporal scalings at $t_j<t<t_{\rm
NR}$ from \citet{Rhoads99} and \citet{SPH99}. In Appendix
\ref{snapshot} we provide expressions for the peak flux and break
frequencies as a function of the physical parameters and solve
them for the physical parameter as a function of the peak flux and
break frequencies for both models {\bf 1} and {\bf 2}. The results
for GRB 030329 are given below.

For GRB 030329, \citet{Berger03} infer $\nu_{sa}\approx 19\;$GHz,
$\nu_m\approx 43\;$GHz and $F_{\nu,{\rm max}}\approx 96\;$mJy at
$t\approx 10\;$days, as well as $p=2.2$. Using Eqs. 4.13-4.16 of
\citet{SE01}, \citet{Berger03} find $E_{\rm iso,52}\approx
0.56\nu_{c,13}^{1/4}$, $n_0\approx 1.8\nu_{c,13}^{3/4}$
$\epsilon_B\approx 0.10\nu_{c,13}^{-5/4}$, $\epsilon_e\approx
0.16\nu_{c,13}^{1/4}$, and using a value of $\theta_j\approx 0.3$
at this time they inferred $E_{51}\approx 0.25$. For the same
values of the spectral parameters and using our {\bf model 1} we
obtain $E_{\rm iso,52}=0.16\nu_{c,13}^{1/4}$,
$E_{51}=0.36\nu_{c,13}^{3/8}$, $n_0=15\nu_{c,13}^{3/4}$,
$\epsilon_B=0.12\nu_{c,13}^{-5/4}$,
$\epsilon_e=0.24\nu_{c,13}^{1/4}$ for $k=0$ and $E_{\rm
iso,52}=0.10\nu_{c,13}^{1/4}$, $E_{51}=0.43\nu_{c,13}^{3/8}$,
$A_*=1.4\nu_{c,13}^{1/2}$, $\epsilon_B=0.034\nu_{c,13}^{-5/4}$,
$\epsilon_e=0.36\nu_{c,13}^{1/4}$ for $k=2$. The implied
values of $E/A$ are shown in Table 1. The differences between our
values and those of \citet{Berger03} arise from differences by
factors of order unity between the coefficients in the expressions
for the peak flux and break frequencies. This typically results in
differences by factors of order unity in the inferred values of
the physical parameters. The difference in the external density
$n$ is relatively large since it contains high powers of
$\nu_{sa}$ and $\nu_m$ \citep{GPS99b} making it more sensitive to
the exact theoretical expressions and observational values of
these frequencies.

For {\rm model 1} and $k=0$ we obtain
$E_{51}/n_0=0.0.024\nu_{c,13}^{-3/8}$ compared to
$E_{51}/n_0=0.14\nu_{c,13}^{-1/2}$ of \citet{Berger03} and
$E_{51}/n_0\sim 0.8$ that we obtain from the fit to the observed
image size (Figure \ref{R_perp_fit}). Because of the large
uncertainty in the value of $n$ that is determined from the
snapshot spectrum, and the large uncertainty in the value of $E/n$
from the fit to the image size, these values are consistent with
each other within their reasonable errors (see Table 1). For {\rm
model 1} and $k=2$ we obtain $E_{51}/A_*=0.31\nu_{c,13}^{-1/8}$
compared to $E_{51}/A_*\approx 0.8$ from the fit to the observed
image size. Here the difference between the two values is smaller,
but the uncertainty on the two values is also smaller (see Table
1). Altogether, the two values are still consistent within their
estimated errors.

For our {\bf model 2} involving a jet with no significant lateral
spreading, the peak flux is suppressed by a factor of
$(t/t_j)^{-(3-k)/(4-k)}$ where $t_j\approx 0.5\;$days and $t/t_j\approx
20$, i.e. a factor of $\approx 0.11$ for $k=0$ and $\approx 0.22$ for
$k=2$. This implies (see appendix \ref{snapshot}) $E_{\rm
iso,52}=4.7\nu_{c,13}^{1/4}$, $E_{51}=0.21\nu_{c,13}^{3/8}$,
$n_0=0.53\nu_{c,13}^{3/4}$, $\epsilon_B=0.37\nu_{c,13}^{-5/4}$,
$\epsilon_e=0.078\nu_{c,13}^{1/4}$ for $k=0$ and $E_{\rm
iso,52}=0.98\nu_{c,13}^{1/4}$, $E_{51}=0.29\nu_{c,13}^{3/8}$,
$A_*=1.4\nu_{c,13}^{1/2}$, $\epsilon_B=0.071\nu_{c,13}^{-5/4}$,
$\epsilon_e=0.17\nu_{c,13}^{1/4}$ for $k=2$. For {\bf model 2} with $k=0$
we get $E_{51}/n_0=0.40\nu_{c,13}^{-3/8}$ compared to $E_{51}/n_0\approx 5$
from the fit to the observed image size.  These two values are consistent
within the large uncertainties on both values (see Table 1).

For {\bf model 2} with $k=2$ we obtain $E_{51}/A_*=0.10\nu_{c,13}^{-1/8}$
compared to $E_{51}/A_*\approx 1.2$ from the fit to the observed image
size. In this case, however, the errors on these two values are relatively
small (see Table 1). This is because: {\it (i)} the image size is linear in
$E/A$ which corresponds to a relatively strong dependence, and therefore
the observed image size can constrain the value of $E/A$ relatively well,
and {\it (ii)} the expression for $E/A$ from the spectrum contains
relatively small powers of the break frequencies and peak flux and thus has
a correspondingly small uncertainty. Therefore, the two values of
$E_{51}/A_*$ are farther apart than is expected from the uncertainty on
these values. Thus, one might say that the data disfavors {\bf model 2}
with $k=2$. It is hard, however, to rule out this model altogether, because
of the uncertainty is the exact expressions for the break frequencies and
peak flux at $t_j<t<t_{\rm NR}$.

\section{Discussion}
\label{dis}

We have analyzed the data on the time-dependent image size of the
radio afterglow of GRB 030329 (Taylor et al. 2004) and constrained
the physical parameters of this explosion. The image size was
measured after the jet break time $t_j$ in the afterglow
lightcurve, where existing theoretical models still have a high
level of uncertainty regarding the jet dynamics. This motivated us
to consider two extreme models for the lateral expansion of the
jet: {\bf model 1}, where there is relativistic lateral expansion
in the local rest frame of the jet at $t_j<t<t_{\rm NR}$, and {\bf
model 2}, with no significant lateral expansion until the
transition time to a non-relativistic expansion $t_{\rm NR}$. We
have tested the predictions of these models against the
observations, for both a uniform ($\rho_{\rm ext}=Ar^{-k}$, with
$k=0$) and a stellar wind ($k=2$) external density profile.

The observational constraints included comparisons between: {\it
(i)} the value of the post-shock energy fraction in relativistic
electrons $\epsilon_e$ that is inferred from the source size and
flux below the self absorption frequency and its value from the
`snapshot' spectrum at $t\approx 10\;$days; {\it (ii)} the value
of $E/A$ that is inferred from the source size and its value from
the `snapshot' spectrum at $t\approx 10\;$days; and {\it (iii)}
the observed temporal evolution of the source size and the
theoretical predictions. We have found that most models pass all
these tests. The only exception is {\bf model 2} with $k=2$,
involving a relativistic jet with little lateral expansion (well
before $t_{\rm NR}$) that is propagating in a stellar wind
external medium, which does poorly on point {\it (ii)} above.

We have found that for a jet with little lateral expansion before
$t_{\rm NR}$ (our {\bf model 2}), the jet would become roughly
spherical only long after $t_{\rm NR}$ (see Eq. \ref{t_sph} and the
discussion around it). This introduces a fast growth in the image size
near $t_{\rm NR}$ for on-axis observers with $\theta_{\rm
obs}<\theta_0$ (see upper panel of Figure \ref{R_perp_t}) that detect
the prompt gamma-ray emission (as in the case of GRB 030329).  For an
observer at $\theta_{\rm obs}\approx 90^\circ$ as would typically be
the case for GRBs that might be found in nearby SNe Ib/c, months to
years after the SN \citep{Pa01,GL03,RM04}, this causes a very slow
increase in the image size near $t_{\rm NR}$ (see lower panel of
Figure \ref{R_perp_t}).

\citet{ONP04} have considered a jet with no lateral spreading,
even at $t\gtrsim t_{\rm NR}$, and concluded that it can be ruled
out for a uniform external density ($k=0$) since it gives $a=1/4$
at $t_j<t<t_{\rm NR}$ which is inconsistent with observations
[recall that in \S \ref{application} we have found that
$a=0.71_{-0.3}^{+0.4}$ ($1\;\sigma$) between 25 and $83\;$days].
In our analysis we have argued that physically one expects lateral
spreading to start around $t_{\rm NR}$, even if it is negligible
at $t\ll t_{\rm NR}$. We have shown that with this more realistic
assumption for the jet dynamics (our model 2) the temporal
evolution of the image size for a uniform external density ($k=0$)
is consistent with observations.

The formalism developed in this paper would be useful for the analysis of
future radio imaging of nearby GRB afterglows.  The forthcoming {\it Swift}
satellite (http://swift.gsfc.nasa.gov/) is likely to discover new GRBs at
low redshifts. Follow-up imaging of their radio jets will constrain their
physical properties and reveal whether the conclusions we derived for GRB
030329 apply more generally to other relativistic explosions.

\acknowledgments

We thank Tsvi Piran, Yonatan Oren and Ehud Nakar for useful discussions
which helped improve the paper. This work was supported by the W.M. Keck
foundation, NSF grant PHY-0070928 (JG), and by NASA through a Chandra
Postdoctoral Fellowship award PF3-40028 (ER-R). It was also supported in
part by NASA grant NAG 5-13292, and by NSF grants AST-0071019, AST-0204514
(A.L.).

\appendix

\newcommand{\rb}[1]{\raisebox{1.5ex}[0pt]{#1}}
\begin{deluxetable}{|l|c|ccclc|}
\tabletypesize{\footnotesize} \tablecaption{Comparing the values
of Physical Parameters derived from different Observables}
\tablewidth{0pt} \tablehead{ \colhead{external} &
\colhead{physical} &
\colhead{observables} &  &  & major source & uncertain by \\
\colhead{density} & \colhead{parameter} & \colhead{being used} &
\rb{\colhead{model 1}} & \rb{\colhead{model 2}} & of uncertainty &
a factor of} \startdata
 &  & $F_\nu(10\,{\rm d})$ & $0.024\nu_{c,13}^{-3/8}$ &
$0.40\nu_{c,13}^{-3/8}$ &
$\propto\nu_{sa}^{-15/4}\nu_m^{-15/8}F_{\nu,{\rm max}}^{9/4}$ &
$\sim 10-100$
\\
\rb{$k=0$} & \rb{$E_{51}/n_0$} & $R_\perp(t)$ & 0.8 & 5 & $\propto
R_\perp^{6\;(3)}$ in model $1\;(2)$  & $\sim 10$ ($\sim 5$)
\\ \hline
 & & $F_\nu(10\,{\rm d})$ & $0.31\nu_{c,13}^{-1/8}$ &
$0.10\nu_{c,13}^{-1/8}$ &
$\propto\nu_{sa}^{-5/4}\nu_m^{-5/8}F_{\nu,{\rm max}}^{3/4}$ &
$\sim 2-3$
\\
\rb{$k=2$} & \rb{$E_{51}/A_*$} & $R_\perp(t)$ & 2 & 1.2 & $\propto
R_\perp^{3\;(1)}$ in model $1\;(2)$ & $\sim 5$ ($\sim 2$)
\\ \hline
 &  &  & 2.4 & 1.5 &
 $R_\perp({\rm obs})$ \& $\Gamma(\theta_*)$ & $\sim 1.3$\\
\rb{$k=0$} & \rb{$\Gamma(25\;{\rm d})$} & \rb{$R_\perp$} &
  2.1 & 2.4  & $R_\perp({\rm obs})$ \& jet model & $\sim 1.1-1.2$ \\
\hline
  &
 &  & 2.4 & 2.4 & $R_\perp({\rm obs})$ \& $\Gamma(\theta_*)$ & $\sim 1.3$\\
\rb{$k=2$} & \rb{$\Gamma(25\;{\rm d})$} & \rb{$R_\perp$} & 2.6 &
2.8 & $R_\perp({\rm obs})$ \& jet  model & $\sim 1.1-1.2$
\\ \hline
 &  &  & 0.023 & 0.10 & $R_\perp$ \& $\Gamma(\theta_*)$ in Eq. \ref{eps_e}
 & $\sim 10$ \\
 $k=0$ & $\epsilon_e$ & \rb{$R_\perp$, $F_{\nu<\nu_{sa}}$} & 0.035 &
 0.024 & $R_\perp$ \& $\Gamma({\rm Fig.\;\ref{R_perp_fit}})$ in Eq. \ref{eps_e} & $\sim 5-10$ \\
 &  & $F_\nu(10\;{\rm d})$ &
$0.24\nu_{c,13}^{1/4}$ & $0.078\nu_{c,13}^{1/4}$ & model \& value
of $\nu_c$ & $\sim 3$
 \\
\hline
  &  &  & 0.023 & 0.023 & $R_\perp$ \& $\Gamma(\theta_*)$ in
  Eq. \ref{eps_e} & $\sim 10$ \\
 $k=2$ & $\epsilon_e$ & \rb{$R_\perp$, $F_{\nu<\nu_{sa}}$} &
0.020 & 0.017 &
   $R_\perp$ \& $\Gamma({\rm Fig.\;\ref{R_perp_fit}})$ in Eq. \ref{eps_e} & $\sim 5-10$ \\
 & &  $F_\nu(10\;{\rm d})$ & $0.36\nu_{c,13}^{1/4}$ &
$0.17\nu_{c,13}^{1/4}$ & model \& value of $\nu_c$ & $\sim 3$

\enddata \tablecomments{Estimates for the physical parameters of GRB
  030329 derived from different observable quantities for different
  models of the jet lateral expansion. The value of $E/A$ is estimated
  from the spectrum at $10\;$days (upper line) and from the fit to the
  observed image size (lower line). The value of $\Gamma(25\;{\rm d})$
  is evaluated both as $\Gamma(\theta_*)$ according to \S
  \ref{self_abs} (upper line) and from the fit to the observed image
  size (lower line). The value of $\epsilon_e$ in first two lines is
  evaluated first using Eq. \ref{eps_e} with the values of
  $\Gamma(25\;{\rm d})$ from the corresponding lines. In the third
  line the value of $\epsilon_e$ is from the spectrum at $10\;$days
  (third line).}

\end{deluxetable}

\section{The Angle $\theta_*$ on the Equal Arrival Time Surface where $R_\perp$ is Located}
\label{theta_*}

The time at which a photon emitted at a lab frame time $t_{\rm
lab}$ and at spherical coordinates $(r,\theta,\phi)$ reaches the
observer is given by
\begin{equation}\label{t}
t=t_{\rm lab}-(R/c)\cos\theta
\end{equation}
and shall be referred to as the observed time, where for
convenience the direction to the observer was chosen to be along
the $z$-axis (i.e. at $\theta=0$). Let the location of a
spherically symmetric shock front (or any other emitting surface
for that matter) be described by $r=R(t_{\rm lab})$ and that of an
axially symmetric shock front by $r=R(t_{\rm lab},\theta)$. We
shall now calculate the angle $\theta_*$ on the equal arrival time
surface (which is defined by $t={\rm const}$) where
$R_\perp=\max(R\sin\theta)$ is located. At this point on the equal
arrival time surface we have
\begin{equation}\label{maximum}
0 =\left(\frac{\partial
R\sin\theta}{\partial\theta}\right)_t=\left(\frac{\partial
R\sin\theta}{\partial\theta}\right)_{t_{\rm
lab}}+\left(\frac{\partial R\sin\theta}{\partial t_{\rm
lab}}\right)_{\theta}\left(\frac{\partial t_{\rm
lab}}{\partial\theta}\right)_{t} =
R(\cos\theta+\tilde{R}_\theta\sin\theta)+\beta_r
c\sin\theta\left(\frac{\partial t_{\rm
lab}}{\partial\theta}\right)_t\ ,
\end{equation}
where we use the notions $(\partial R/\partial t_{\rm
lab})_\theta=\beta_{r}c$ and $\tilde{R}_\theta=(\partial\ln
R/\partial\theta)_{t_{\rm lab}}$. From Eq. (\ref{t}) we have
\begin{equation}\label{dtdth}
0=\left(\frac{\partial t}{\partial\theta}\right)_t=
\frac{R}{c}(\sin\theta-\tilde{R}_\theta\cos\theta)+
(1-\beta_r\cos\theta) \left(\frac{\partial t_{\rm
lab}}{\partial\theta}\right)_t\ ,
\end{equation}
so that
\begin{equation}\label{dtdth2}
\left(\frac{\partial t_{\rm
lab}}{\partial\theta}\right)_t=\frac{R}{c}
\left(\frac{\tilde{R}_\theta\cos\theta-\sin\theta}{1-\beta_r\cos\theta}\right)\
.
\end{equation}
Substituting Eq. (\ref{dtdth2}) into Eq. (\ref{maximum}) we obtain
\begin{equation}\label{th*}
\cos\theta=\beta_r-\tilde{R}_\theta\sin\theta=\frac{1}{c}\left(\frac{\partial
R}{\partial t_{\rm
lab}}\right)_{\theta}-\frac{\sin\theta}{R}\left(\frac{\partial
R}{\partial\theta}\right)_{t_{\rm lab}}\ .
\end{equation}
For a spherically symmetric shock $\tilde{R}_\theta=0$ and
$\cos\theta_*=\beta_r(\theta_*)=\beta_*$, where in this case
$\beta_r$ is the shock velocity at the point on the equal arrival
time surface were $\theta=\theta_*$ and $R_\perp$ is located. For
a shock with axial symmetry we have
\begin{equation}\label{th*2}
\cos\theta_*=\frac{\beta_r-\tilde{R}_\theta
\sqrt{1-\beta_r^2+\tilde{R}_\theta^2}}{1+\tilde{R}_\theta^2}\ ,
\end{equation}
and
\begin{equation}\label{beta_r}
\beta_r=\beta_*\sqrt{1+\tilde{R}_\theta^2}\ ,
\end{equation}
where $\beta_*$ is the shock velocity component normal to the
shock front in the rest frame of the upstream medium, which is the
one that enters into the shock jump conditions \citep{KG03}.

\section{The Apparent Velocity}
\label{app_vel}

The apparent velocity, $\beta_{\rm ap}=[(1+z)/c](dR_\perp/dt)$,
for a point source moving with a velocity $\beta$ at an angle
$\theta$ from our line-of-sight is
\begin{equation}\label{beta_ap}
\beta_{\rm ap}=\frac{\beta\sin\theta}{1-\beta\cos\theta}\ .
\end{equation}
For a spherical shock front moving at a constant velocity
$\beta_{\rm sh}$, $R_\perp$ is located at a constant angle
$\theta_*$ which satisfies $\cos\theta_*=\beta_*=\beta_{\rm
sh}={\rm const}$ (according to Eq. \ref{th*}) so that the apparent
velocity of the edge of the observed image is simply given by
substituting $\cos\theta_*=\beta_*$ in Eq. (\ref{beta_ap}). This
gives
\begin{equation}\label{beta_ap2}
\beta_{\rm ap}=\Gamma_*\beta_*=\sqrt{\Gamma_*^2-1}\ .
\end{equation}
We shall now show that this result holds for any spherically
symmetric shock front. At $t+dt$ we have
\begin{equation}\label{dth}
\theta_*(t+dt)=\theta_*(t)+d\theta_*\quad\quad,\quad\quad
\beta_*(t+dt)=\beta_*(t)+d\beta_*\quad\quad,\quad\quad
d\beta_*=d\cos\theta_*\propto dt\ ,
\end{equation}
and since Eq. (\ref{beta_ap2}) holds for a sphere moving at a
constant velocity, we have
\begin{equation}\label{R_p}
\left[R\sin\theta\right](t+dt,\theta_*)=
R_\perp(t)+\Gamma_*(t)\beta_*(t)cdt+{\mathcal
O}(dt^2)\ .
\end{equation}
Now, since $R_\perp$ is located where $(\partial
R\sin\theta/\partial\theta)_t=0$ then
\begin{equation}\label{R_p2}
R_\perp(t+dt)=\left[R\sin\theta\right](t+dt,\theta_*+d\theta_*)=
\left[R\sin\theta\right](t+dt,\theta_*)+{\mathcal O}(dt^2) =
R_\perp(t)+\Gamma_*(t)\beta_*(t)cdt+{\mathcal O}(dt^2)\ ,
\end{equation}
and therefore  Eq. (\ref{beta_ap2}) holds for any spherically
symmetric shock front.

Finally, for an axially symmetric shock front, we obtain based on similar
considerations as in the spherical case
\begin{equation}\label{2D}
\beta_{\rm ap}=\frac{\beta_r\sin\theta_*}{1-\beta_r\cos\theta_*}\
,
\end{equation}
where $\theta_*$ and $\beta_r$ are given by Eqs. \ref{th*2} and
\ref{beta_r}, respectively.

\section{Solving for the Physical Parameters from a `Snapshot'
Spectrum at $t>t_j$} \label{snapshot}

The most common ordering of the spectral break frequencies at
$t_j<t<t_{\rm NR}$ is $\nu_{sa}<\nu_m<\nu_c$, for which we obtain
\begin{eqnarray}
\nu_{sa} &=& 2.08\times 10^9\frac{(p-1)^{8/5}}{(p-2)(3p+2)^{3/5}}
(1+z)^{-4/5}\epsilon_e^{-1}\epsilon_B^{1/5}n_0^{8/15}E_{51}^{4/15}t_{\rm
days}^{-1/5}\;{\rm Hz}\ ,
\\
\nu_m &=& 1.35\times
10^{16}\left(\frac{p-2}{p-1}\right)^2(p-0.67)(1+z)
\epsilon_e^{2}\epsilon_B^{1/2}n_0^{-1/6}E_{51}^{2/3}t_{\rm
days}^{-2}\;{\rm Hz}\ ,
\\
\nu_c &=& 1.75\times 10^{13}(p-0.46)e^{-1.16p}(1+z)^{-1}
\epsilon_B^{-3/2}n_0^{-5/6}E_{51}^{-2/3}(1+Y)^{-2}\;{\rm Hz}\ ,
\\
F_{\nu,{\rm max}} &=&
131(p+0.14)(1+z)^2\epsilon_B^{1/2}n_0^{1/6}E_{51}^{4/3}t_{\rm
days}^{-1}D_{L,28}^{-2}\;{\rm mJy}\ ,
\end{eqnarray}
for a uniform external medium ($k=0$), and
\begin{eqnarray}
\nu_{sa} &=& 3.85\times 10^9\frac{(p-1)^{8/5}}{(p-2)(3p+2)^{3/5}}
(1+z)^{-4/5}\epsilon_e^{-1}\epsilon_B^{1/5}A_*^{8/5}E_{51}^{-4/5}t_{\rm
days}^{-1/5}\;{\rm Hz}\ ,
\\
\nu_m &=& 1.05\times
10^{16}\left(\frac{p-2}{p-1}\right)^2(p-0.69)(1+z)
\epsilon_e^{2}\epsilon_B^{1/2}A_*^{-1/2}E_{51}t_{\rm
days}^{-2}\;{\rm Hz}\ ,
\\
\nu_c &=& 1.15\times 10^{11}(3.45-p)e^{0.45p}(1+z)^{-1}
\epsilon_B^{-3/2}A_*^{-5/2}E_{51}(1+Y)^{-2}\;{\rm Hz}\ ,
\\
F_{\nu,{\rm max}} &=&
201(p+0.12)(1+z)^2\epsilon_B^{1/2}A_*^{1/2}E_{51}t_{\rm
days}^{-1}D_{L,28}^{-2}\;{\rm mJy}\ ,
\end{eqnarray}
for a stellar wind environment ($k=2$), where $Y$ is the Compton
y-parameter, $A_*=A/(5\times 10^{11}\;{\rm gr\;cm^{-1}})$, $t_{\rm
days}=t/(1\;{\rm day})$, $\epsilon_B$ is the fraction of the
internal energy behind the shock in the magnetic field, and
$Q_x\equiv Q/(10^x\times{\rm the\ c.g.s.\ units\ of\ }Q)$. The
emission depends only on the true energy in the jet, $E$, and does
not depend on its initial half-opening angle $\theta_0$, since at
$t>t_j$ (or equivalently when $\Gamma$ drope below
$\theta_0^{-1}$) the dynamics become independent of $\theta_0$,
i.e. the jet begins to expand sideways exponentially with radius
in a self similar manner that is independent of $\theta_0$
\citep{G02}. Solving the above sets of equations for the physical
parameters yields
\begin{eqnarray}
E_{\rm iso,52} &=&
0.104\frac{f_0(p)}{f_0(2.2)}\nu_{a,9}^{-5/6}\nu_{m,13}^{-5/12}
\nu_{c,14}^{1/4}\left(\frac{F_{\nu,{\rm max}}}{1\;\rm
mJy}\right)^{3/2}t_{\rm days}^{-1/2}
(1+z)^{-2}D_{L,28}^{3}(1+Y)^{1/2}\ ,
\\ \label{Ek0}
E_{51} &=& 0.0136\frac{g_{0,E}(p)}{g_{0,E}(2.2)}
\nu_{a,9}^{5/12}\nu_{m,13}^{5/24}\nu_{c,14}^{3/8}
\left(\frac{F_{\nu,{\rm max}}}{1\;\rm mJy}\right)^{3/4}t_{\rm
days}^{5/4}(1+z)^{-1}D_{L,28}^{3/2}(1+Y)^{3/4}\ ,
\\
n_0 &=& 0.0714\frac{g_{n}(p)}{g_{n}(2.2)}
\nu_{a,9}^{25/6}\nu_{m,13}^{25/12}\nu_{c,14}^{3/4}
\left(\frac{F_{\nu,{\rm max}}}{1\;\rm mJy}\right)^{-3/2}t_{\rm
days}^{7/2}(1+z)^{5}D_{L,28}^{-3}(1+Y)^{3/2}\ ,
\\
\epsilon_B &=& 2.42\frac{g_{0,B}(p)}{g_{0,B}(2.2)}
\nu_{a,9}^{-5/2}\nu_{m,13}^{-5/4}\nu_{c,14}^{-5/4}
\left(\frac{F_{\nu,{\rm max}}}{1\;\rm mJy}\right)^{1/2}t_{\rm
days}^{-5/2}(1+z)^{-3}D_{L,28}(1+Y)^{-5/2}\ ,
\\
\epsilon_e &=& 0.355\frac{g_{0,e}(p)}{g_{0,e}(2.2)}
\nu_{a,9}^{5/6}\nu_{m,13}^{11/12}\nu_{c,14}^{1/4}
\left(\frac{F_{\nu,{\rm max}}}{1\;\rm mJy}\right)^{-1/2}t_{\rm
days}^{3/2}(1+z)D_{L,28}^{-1}(1+Y)^{1/2}\ ,
\end{eqnarray}
for a uniform density, where
$f_0(p)=e^{0.29p}(p-1)^{1/2}(3p+2)^{-1/2}(p-0.67)^{5/12}
(p-0.46)^{-1/4}(p+0.14)^{-3/2}$,
$g_{0,E}(p)=e^{0.435p}(p-1)^{-1/4}(3p+2)^{1/4}(p-0.67)^{-5/24}
(p-0.46)^{-3/8}(p+0.14)^{-3/4}$,
$g_{n}(p)=e^{0.87p}(p-1)^{-5/2}(3p+2)^{5/2}(p-0.67)^{-25/12}
(p-0.46)^{-3/4}(p+0.14)^{3/2}$,
$g_{0,B}(p)=e^{-1.45p}(p-1)^{3/2}(3p+2)^{-3/2}(p-0.67)^{5/4}
(p-0.46)^{5/4}(p+0.14)^{-1/2}$,
$g_{0,e}(p)=e^{0.29p}(p-2)^{-1}(p-1)^{1/2}(3p+2)^{1/2}(p-0.67)^{-11/12}
(p-0.46)^{-1/4}(p+0.14)^{1/2}$. For a stellar wind environment we
find
\begin{eqnarray}
E_{\rm iso,52} &=&
0.0674\frac{f_2(p)}{f_2(2.2)}\nu_{a,9}^{-5/6}\nu_{m,13}^{-5/12}\nu_{c,14}^{1/4}
\left(\frac{F_{\nu,{\rm max}}}{1\;\rm mJy}\right)^{3/2}t_{\rm
days}^{-1/2} (1+z)^{-2}D_{L,28}^{3}(1+Y)^{1/2}\ ,
\\ \label{Ek2}
E_{51} &=& 0.0161\frac{g_{2,E}(p)}{g_{2,E}(2.2)}
\nu_{a,9}^{5/12}\nu_{m,13}^{5/24}\nu_{c,14}^{3/8}
\left(\frac{F_{\nu,{\rm max}}}{1\;\rm mJy}\right)^{3/4}t_{\rm
days}^{5/4}(1+z)^{-1}D_{L,28}^{3/2}(1+Y)^{3/4}\ ,
\\
A_* &=& 0.0262\frac{g_{A}(p)}{g_{A}(2.2)}
\nu_{a,9}^{5/3}\nu_{m,13}^{5/6}\nu_{c,14}^{1/2}t_{\rm
days}^{2}(1+z)(1+Y)\ ,
\\
\epsilon_B &=& 0.680\frac{g_{2,B}(p)}{g_{2,B}(2.2)}
\nu_{a,9}^{-5/2}\nu_{m,13}^{-5/4}\nu_{c,14}^{-5/4}
\left(\frac{F_{\nu,{\rm max}}}{1\;\rm mJy}\right)^{1/2}t_{\rm
days}^{-5/2}(1+z)^{-3}D_{L,28}(1+Y)^{-5/2}\ ,
\\
\epsilon_e &=& 0.526\frac{g_{2,e}(p)}{g_{2,e}(2.2)}
\nu_{a,9}^{5/6}\nu_{m,13}^{11/12}\nu_{c,14}^{1/4}
\left(\frac{F_{\nu,{\rm max}}}{1\;\rm mJy}\right)^{-1/2}t_{\rm
days}^{3/2}(1+z)D_{L,28}^{-1}(1+Y)^{1/2}\ ,
\end{eqnarray}
where $f_2(p)=e^{-0.113p}(p-1)^{1/2}(3p+2)^{-1/2}
(p-0.69)^{5/12}(3.45-p)^{-1/4}(p+0.12)^{-3/2}$,
$g_{2,E}(p)=e^{-0.169p}(p-1)^{-1/4}(3p+2)^{1/4}
(p-0.69)^{5/4}(3.45-p)^{5/4}(p+0.12)^{3/4}$,
$g_{A}(p)=e^{-0.225p}(p-1)^{-1}(3p+2)(p-0.69)^{-5/6}(3.45-p)^{-1/2}$,
$g_{2,B}(p)=e^{0.563p}(p-1)^{3/2}(3p+2)^{-3/2}
(p-0.69)^{5/4}(3.45-p)^{5/4}(p+0.12)^{-1/2}$,
$g_{2,e}(p)=e^{-0.113p}(p-2)^{-1}(p-1)^{1/2}(3p+2)^{1/2}
(p-0.69)^{-11/12}(3.45-p)^{-1/4}(p+0.12)^{1/2}$.

As was pointed out by \citet{SE01}, the expressions for the
physical parameters that are derived from the instantaneous
(`snapshot') spectrum do not depend on the external density
profile (i.e. on the value of $k$ in our case), up to factors of
order unity. This is because the instantaneous spectrum samples
only the instantaneous external density just in front of the
afterglow shock, $n_{\rm ext}(r)$. The expression for the external
density $n$ for a uniform medium ($k=0$) represents the density
just in front of the shock for a general density profile that
varies smoothly and gradually with radius, $n \longleftrightarrow
n_{\rm ext}(r)$, where in our case $n_{\rm ext}=Ar^{-k}/m_p$.
However, for a non-uniform density $n_{\rm ext}$ changes with
radius and therefore with time. In our case, we assume the
functional form of $n_{\rm ext}(r)$ is known (i.e. we fix the
value of $k$) and express the density normalization $A$ as a
function of the instantaneous values of the peak flux and break
frequencies.

We note that the expressions for the physical parameters at
$t_j<t<t_{\rm NR}$ are identical to those at $t<t_j$. This is
because we assume that the jet is uniform within a half-opening
angle $\theta_j\approx\Gamma^{-1}$, and therefore its emission is
practically indistinguishable from that of a spherical blast wave
with the same Lorentz factor $\Gamma$ and radius $R$, or
equivalently\footnote{This is since $\Gamma$ and $R$ are functions
of $E_{\rm iso}$, $n_{\rm ext}=\rho_{\rm ext}/m_p$ and $t$:
$E_{\rm iso}\sim\Gamma^2 R^2\rho_{\rm ext}c^2$ and $t\sim
R/c\Gamma^2$ so that $R\sim(E_{\rm iso}t/\rho_{\rm ext}c)^{1/4}$
and $\Gamma\sim(E_{\rm iso}/\rho_{\rm ext}c^5t^3)^{1/8}$.} the
same isotropic equivalent energy $E_{\rm iso}$ (which for a
spherical blast wave is equal to the true energy, and for a {\bf
model 1} jet is $E_{\rm iso}\approx(2/\theta_j^2)E\approx
2\Gamma^2 E$) and observed time $t$ (for the same values of
$n_{\rm ext}$, $\epsilon_e$, $\epsilon_B$ and $p$).

At $t<t_j$, $E_{\rm iso}={\rm const}$ and is the more interesting
physical quantity, while $E$ in Eqs. (\ref{Ek0}) and (\ref{Ek2})
represents the energy within an angle of $\Gamma^{-1}$ around our
line-of-sight which has no special physical significance at this
stage. At $t_j<t<t_{\rm NR}$, however, the situation is reversed
and $E={\rm const}$ represents the true kinetic energy of the jet,
and is therefore of great interest, while $E_{\rm iso}\approx
2\Gamma^2 E$ decreases with time and is no longer a very
interesting physical quantity.

For {\bf model 2}, the jet continues to evolve as if it were part
of a spherical blast wave with the same $E_{\rm iso}$ until
$t_{\rm NR}(E_{\rm iso})$, and $E_{\rm
iso}\approx(2/\theta_0^2)E={\rm const}$. Therefore, the emission
at $t_j<t<t_{\rm NR}$ is the same as from a spherical blast wave
with the same $E_{\rm iso}$, except for the peak flux $F_{\nu,{\rm
max}}$ which is suppressed by a factor of
$\sim(\theta_0\Gamma)^2\approx(t/t_j)^{-(3-k)/(4-k)}$. Hence, the
above equations for the physical parameters may still be used in
this case with the substitution $F_{\nu,{\rm max}}\longrightarrow
F_{\nu,{\rm max}}(t/t_j)^{(3-k)/(4-k)}$. In addition to this, in
order to obtain the true energy in the jet, the expression for $E$
(Eqs. \ref{Ek0} and \ref{Ek2}) should be multiplied by
$(t/t_j)^{-(3-k)/(4-k)}$, which is the fraction of the area within
an angle of $\Gamma^{-1}$ around the line-of-sight which is
occupied by the jet.

\end{document}